\documentclass[a4paper,10pt]{aastex62}

\hypersetup{linkcolor=red,citecolor=blue,filecolor=cyan,urlcolor=magenta}
\usepackage{natbib}
\usepackage{amsthm}
\usepackage{amssymb}
\usepackage{amsmath}
\usepackage{bm}
\usepackage[utf8]{inputenc}
\usepackage{graphicx}
\usepackage[caption=false]{subfig}
\usepackage{soul}
\usepackage{dblfloatfix}

\shorttitle{PIC simulation of Whistler Heat Flux Instabilities in the near-Sun Solar Wind}
\shortauthors{Micera et al.}

\begin{document}

\title{Particle-In-Cell simulation of whistler heat flux instabilities in the solar wind: heat flux regulation and electron halo formation.}

\correspondingauthor{A. Micera}
\email{alfredo.micera@oma.be}

\author[0000-0001-9293-174X]{A. Micera}
\affil{Solar-Terrestrial Centre of Excellence - SIDC, Royal Observatory of Belgium, Brussels, Belgium.}
\affil{Centre for Mathematical Plasma Astrophysics, KU Leuven, Leuven, Belgium.}

\author[0000-0002-2542-9810]{A. N. Zhukov}
\affiliation{Solar-Terrestrial Centre of Excellence - SIDC, Royal Observatory of Belgium, Brussels, Belgium.}
\affiliation{Skobeltsyn Institute of Nuclear Physics, Moscow State University, Moscow, Russia.}

\author[0000-0003-3223-1498]{R.~A. L\'{o}pez}
\affiliation{Departamento de F\'{i}sica, Universidad de Santiago de Chile, Santiago, Chile}

\author[0000-0002-5782-0013]{M. E. Innocenti}
\affiliation{Centre for Mathematical Plasma Astrophysics, KU Leuven, Leuven, Belgium.}
 
\author[0000-0002-8508-5466]{M. Lazar}
\affiliation{Centre for Mathematical Plasma Astrophysics, KU Leuven, Leuven, Belgium.}
\affiliation{Institut f\"{u}r Theoretische Physik, Ruhr-Universität Bochum, Bochum, Germany}

\author[0000-0003-1970-6794]{E. Boella}
\affiliation{Physics Department, Lancaster University, Lancaster, UK.}
\affiliation{Cockcroft Institute, Daresbury Laboratory, Warrington, UK.}

\author[0000-0002-3123-4024]{G. Lapenta}
\affiliation{Centre for Mathematical Plasma Astrophysics, KU Leuven, Leuven, Belgium.}

\begin{abstract}
We present results of two-dimensional fully kinetic Particle-In-Cell simulation in order to shed light on the role of whistler waves in the scattering of strahl electrons and in the heat flux regulation in the solar wind.
We model the electron velocity distribution \textcolor{black}{function as initially} composed of core and strahl populations as typically encountered in the near-Sun solar wind as observed by Parker Solar Probe. We demonstrate that, \textcolor{black}{as a consequence of the evolution of the electron velocity distribution function}, two branches of the whistler heat flux instability can be excited, which can drive whistler waves propagating in the direction oblique or parallel to the background magnetic field. 
First, oblique whistler waves induce pitch-angle scattering of strahl electrons, \textcolor{black}{towards higher perpendicular velocities. This leads to the broadening of the strahl pitch angle distribution and hence to the formation of a halo-like population at the expense of the strahl. Later on, the electron velocity distribution function experiences the effect of parallel whistler waves, which contributes to the redistribution of the particles scattered in the perpendicular direction into a more symmetric halo, in agreement with observations. Simulation results show a remarkable agreement with the linear theory of the oblique whistler heat flux instability.
The process is accompanied by a significant decrease of the heat flux carried by the strahl population.}
\end{abstract}

\keywords{Plasma Astrophysics --- Solar wind --- Space plasmas}

\section{Introduction}\label{sec.0}

Electron velocity distribution functions (VDFs) in the solar wind often consist of three components. At lower energies, there is a dense, thermal and almost isotropic core, while at higher energies, a suprathermal halo population distributed at all pitch angles and a suprathermal field-aligned strahl can be observed \citep{Feldman1975, Pilipp1987, Gosling2001,Salem2003, Stverak_2009,Halekas2020}.

This peculiar nonthermal structure of the electron VDF carries an important amount of heat flux in the solar wind \citep{Marsch2006}.
In situ observations made at 1 au \citep{Feldman_1977,Bale_2013} and further away from the Sun \citep{Scime1994} show that the heat flux carried by the solar wind is suppressed below the values provided by collisional models \citep{Spitzer_1953}. Since suprathermal electron populations are not much affected by Coulomb collisions (the Coulomb cross section strongly depends on the particle velocities), a wide variety of kinetic instabilities can be responsible for shaping the electron VDF, reducing the skewness of suprathermal features in the distribution, and hence regulating the heat flux \citep[e.g.][]{Gary1975, Scime1994, Gary1994, Lazar2011a, Roberg-Clark2019}. 

Recent observations by the Parker Solar Probe (PSP) mission have shown that in the vicinity of the Sun the halo fractional density represents only a very small percentage of the total electron density, while the strahl is more pronounced \citep{Halekas2020,Bercic2020}. Furthermore, while the halo fractional density increases with the heliocentric distance, the strahl fractional density decreases \citep{Hammond1996,Maksimovic2005,Pagel2007, Stverak_2009,Gurgiolo2012,Anderson2012,Bercic2019}. These observations suggest that the formation of the halo population takes place from pitch-angle scattering of the strahl as the solar wind expands \citep{Pierrard_1996, Maksimovic1997, Landi2003, Boldyrev2019,Tang2020}.

Since deviations from the thermal distribution are more pronounced in the weakly collisional fast solar wind \citep{Ogilvie,PhillipsGosling1990,Landi2003}, and considering that the major part of the heat flux in the fast solar wind is carried by the strahl electrons \citep{Rosenbauer1977,Pilipp1987}, the heat flux suppression and the density exchange between the high-energy components of the electron VDF are two related mechanisms that are fundamental to understand the dynamic of the solar wind at the early stages of its expansion. One of the models often invoked in this respect assumes a fundamental role of whistler-mode waves, generated by electrons-driven instabilities.  
Indeed, in the presence of counter-streaming electron populations, parallel and oblique (with respect to the interplanetary magnetic field) whistler-like fluctuations can be excited by self-generated microinstabilities \citep{Gary1975}. Observations confirm the existence of quasi-parallel whistler fluctuations in the solar wind at 1 au \citep{Lacombe2014} and in the pristine solar wind \citep{Tong2019}. Moreover, recent PSP data, collected in the inner heliosphere, revealed the existence of whistler-mode waves with a polarization range that goes from parallel to highly oblique \citep{Malaspina2020,Agapitov2020,Mozer2020,Cattell2020}. 

\textcolor{black}{It has long been debated whether collisional or collisionless processes are mainly responsible for the heat flux reduction in the solar wind \citep[e.g.][]{Marsch2006}. \citet{Halekas2020} analyzed the first data of the PSP mission and found a better correlation of the strahl and halo fractional densities with the electron core plasma $\beta$, with $\beta$ being the ratio of the thermal pressure to the magnetic pressure, rather than with the collisional age. This suggests that in the near-Sun solar wind the regulation of the non-thermal features of the electron VDF and hence of the electron heat flux may be carried out predominantly by wave-particle interactions rather than by collisions.}

One-dimensional (1D) Particle-In-Cell (PIC) simulations aimed at reproducing the interactions between electron VDFs and parallel whistler waves, generated by whistler heat flux instability (WHFI) in solar wind conditions, have recently been performed \citep[e.g.][]{Lopez2019,Kuzichev2019}. 
Furthermore, theoretical studies on the effects of highly oblique whistler waves in suppressing the electron heat flux and in transferring the strahl electrons into the halo have been carried out \citep{Vasko2019, Verscharen_2019}.\\
\citet{Lopez2020} explored the various possible heat flux related instabilities and their relevance depending on the solar wind conditions. \citet{Innocenti2020} have shown via fully kinetic expanding box model simulations \citep{Innocentietal2019, Innocenti2019} that the solar wind expansion can trigger or modify the evolution of kinetic microinstabilities that can in turn affect the heat flux regulation. However, it remains an open question whether whistler heat flux instabilities and the resulting wave fluctuations can correctly describe the reasons why the strahl fractional density decreases with heliocentric distance and whether they can explain the low level of electron heat flux inconsistent with the Spitzer-Härm predictions \citep{Spitzer_1953}. 

Here, we present kinetic simulations of the full spectrum of whistler-like fluctuations, self-generated by two drifting electron populations, without seeding any instability and without imposing a temperature gradient.
We initialize core-strahl electron VDFs, characteristic of the near-Sun solar wind, and we address, via a non-linear analysis, the process responsible for reducing the strahl drift velocity, \textcolor{black}{for the generation of the electron halo population from pitch-angle scattering of strahl electrons,} and for regulating the electron heat flux. 

This letter is organized as follows. Section \ref{sec.2} illustrates the simulation setup. Section \ref{sec.3} \textcolor{black}{reports the first 2D simulation of the whistler heat flux instabilities triggered by a realistic solar wind electron VDF}.
Section \ref{sec.5} presents a discussion of the simulation results and reports the conclusions.

\section{Setup of the PIC simulation}\label{sec.2}

In order to provide a complete picture of the non-linear interaction between suprathermal elecrons and whistler waves, and to determine the effects of those waves on the heat flux carried by the solar wind, we perform \textcolor{black}{a} two-dimensional (2D) full PIC simulation using the semi-implicit code, iPIC3D~\citep{Markidis2010}.

We model a collisionless plasma with initially uniform background magnetic field. The plasma and magnetic field parameters correspond to those measured by PSP during its first perihelion by the SWEAP \citep{Kasper2019} and FIELDS \citep{Bale2019} instruments, as reported by \citet{Halekas2020}. The magnetic field is directed along the $x$-axis, $\bm{B}_0=B_0 \hat{e}_x$. Its magnitude $B_0 = 60$~nT  is such that $v_{Ae}/c = 0.01$, with $v_{Ae} = B_0/ \sqrt{4 \pi n_e m_e}$ being the electron Alfv{\'e}n speed, $n_e$ is the electron number density, $m_e$ is the electron mass and $c$ is the speed of the light in vacuum. 
The plasma is composed of core (subscript c) and strahl (subscript s) electrons, and ions (subscript i), assumed to be only protons, with real mass ratio $\mu =~m_i / m_e =~1836$. We use $2048$ particles per cell per species. The plasma satisfies the quasi-neutrality condition: 
$n_{i} = n_{e} = n_{c} + n_{s}$, where $n_{i}$, $n_c = 0.95 ~n_e$ and $n_s = 0.05 ~n_e$ are the ion, electron core and electron strahl densities, respectively.
The electron core is characterized by a sunward drift which balances the current carried by the strahl \citep{Feldman1975, Scime1994}:
$n_{c}\; u_{c} + n_{s}\; u_{s}\; = 0$, with $u_{c}(u_s)$ electron core (strahl) drift velocity. \textcolor{black}{We select an initial strahl drift velocity $u_s = 3.1 ~v_{Ae}$ and hence $u_c = -0.16 ~v_{Ae}$ in order to satisfy the zero net-current condition.}

The initial number density is $n_e = 350$~cm$^{-3}$. \textcolor{black}{The temperature of the core is} \textcolor{black}{$k_B T_c = 43$ ~eV}, where $k_{B}$ is the Boltzmann constant, such that \textcolor{black}{$\beta_{c} = 1.6$}, with $\beta_{j} = 8 \pi n_j k_B T_{j} / B_0^2$. Regarding the ions, we assume that their drift velocity is zero ($u_i = 0$) and $\beta_{i} = 2$. \textcolor{black}{Core electrons and ions} are assumed initially isotropic ($T_{j,\perp}/T_{j,\parallel} = 1$) \textcolor{black}{and Maxwellian}. 

\textcolor{black}{For the strahl we adopt a temperature in the direction parallel to the background magnetic field of $k_B T_{s, \parallel} = 179$ ~eV, a temperature anisotropy $T_{s, \parallel} / T_{s, \perp} =2$ (to take into account the limited angular extent of the strahl in PSP observations \citep{Bercic2020}, and the adiabatic focusing of the strahl in the expanding solar wind). The initial VDF is thus a drifting bi-Maxwellian:}


\begin{equation}
     \begin{aligned}
f_j (v_{\parallel}, v_{\perp}, t=0) = &\frac{1}{(2 \pi)^{3/2}~ w^2_{\perp j} w_{\parallel j}} \exp \left(-\frac{v_{\perp}^2}{2 w_{\perp j}^2} - \frac{(v_{\parallel} - u_j)^2}{2 w_{\parallel j}^2} \right), 
    \label{sss}   
     \end{aligned}     
\end{equation}

with $w_j = \sqrt{k_B T_j / m_j}$ being the thermal velocity of the species $j$, while $\perp$ and $\parallel$ denote directions perpendicular and parallel to the background magnetic field, $\bm{B}_0$, \textcolor{black}{with $v_{\parallel} \equiv v_x$.} 
\textcolor{black}{The use of a drifting-Maxwellian or a drifting-bi-Maxwellian model to describe the core and the strahl, respectively, is motivated by the PSP observations of near-Maxwellian VDFs, for both electron populations, close to the Sun \citep{Halekas2020, Bercic2020}. The resulting initial VDF is shown in Figure~\ref{fig.1}(a).} 

With these parameters, we obtain $\omega_{pe} / \Omega_{ce} = 100$, with $\omega_{pe} = \sqrt{4 \pi e^2 n_e / m_e}$ and $\Omega_{ce} = e B_0 / m_e\, c$ being the plasma and the cyclotron frequencies for the electrons, respectively, and $e$ is the elementary charge.

A square simulation box with length $L = 8\, d_i$ has been employed, where $d_i = c / \omega_{pi}$ is the ion inertial length, with $\omega_{pi} = \sqrt{4 \pi e^2 n_i / m_i}$ being the plasma frequency for the ions. A cell size $\Delta x = \Delta y = 0.01\, d_i$ and a temporal step $\Delta t = 0.05\, \omega_{pi}^{-1}$ have been chosen. \textcolor{black}{The ion cyclotron frequency is $\Omega_{ci} = e B_0 / m_i\, c = 0.000233 ~\omega_{pi}$.} 

We note that simulations performed with different resolutions and numbers of particles per cell yield similar results, confirming code convergence.

\section{{Simulation Results}}\label{sec.3}

Various types of heat-flux related instabilities can be triggered in the presence of two counter-streaming populations of electrons, according to the initial parameters. Namely, these are the quasi-parallel WHFI~\citep{Gary1975}, the firehose heat-flux instability (FHFI)~\citep{Shaaban2018MN}, the oblique whistler heat-flux instability (O-WHFI)~\citep{Lopez2020} or electrostatic instabilities of the electron acoustic or electron beam modes~\citep{Gary1978}. For the plasma parameters employed in our simulation, the system is initially subject to the oblique whistler heat-flux instability. This is a right-hand polarized mode with maximum growth rate at oblique angle of propagation.
\textcolor{black}{This instability has gained renewed attention due to recent PSP observations of oblique whistler waves in the solar wind near the Sun ~\citep[e.g.][]{Agapitov2020,Cattell2020}.}

\begin{figure*}
 \centering
 \includegraphics[width=0.75\textwidth]{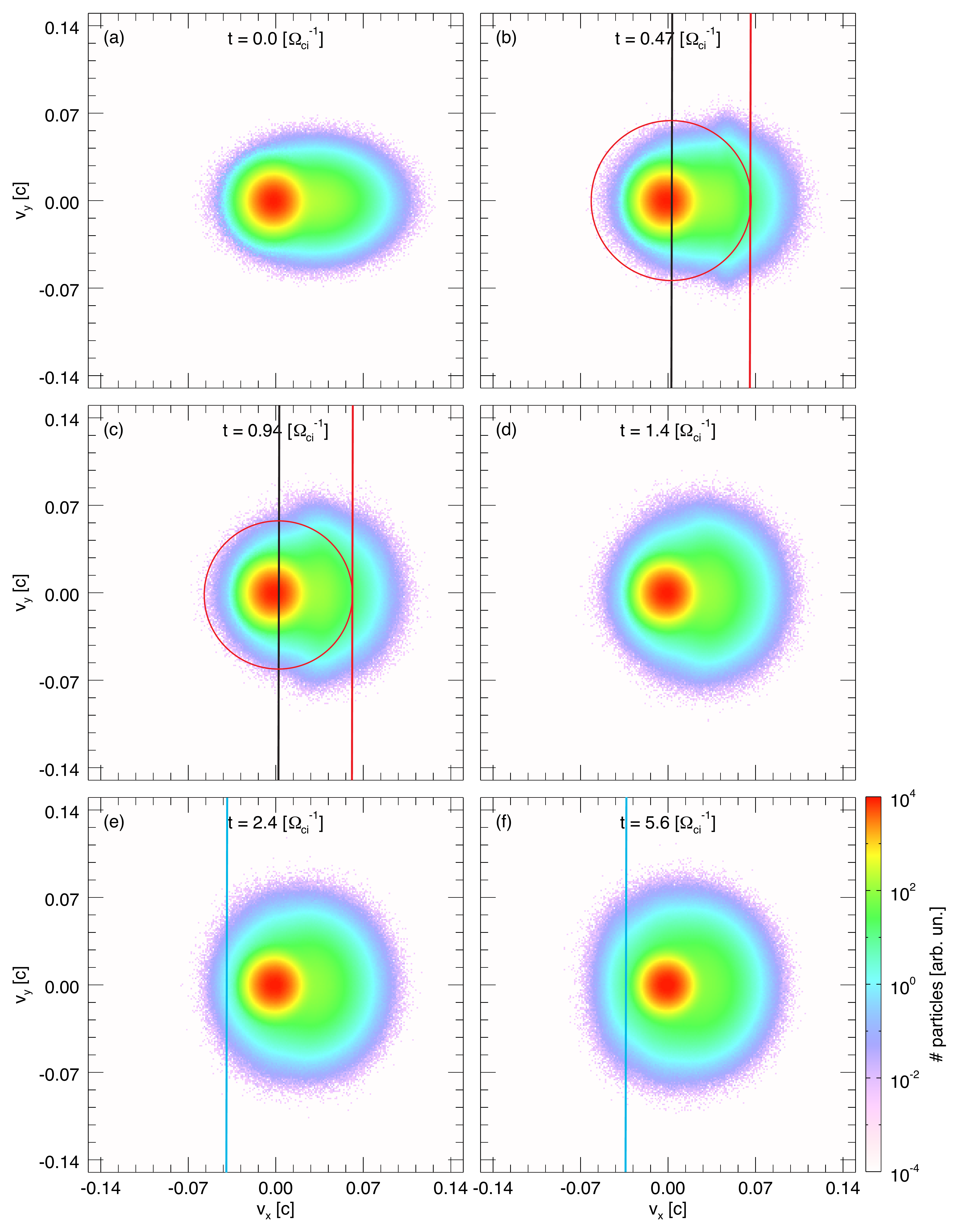}
\caption{Electron VDF $f_e = f(v_x, v_y)$ at $t_0=0$ (initial distribution, panel (a)), $t_1 = 0.47$ (linear stage of O-WHFI, panel (b)), $t_2 = 0.94$ (O-WHFI relaxation, panel (c)), $t = 1.4$ (O-WHFI saturation, panel (d)), $t = 2.4$ (quasi-parallel WHFI, panel (e)) and $t_{end} = 5.6$ (final stage, panel (f)). The time is in units of $\Omega_{ci}^{-1}$. Black and red vertical lines in panels (b) and (c) indicate $v_{\parallel}$ values at which $n=0$ and $n=1$ resonances of oblique whistler waves are expected, while circles show electrons diffusion paths due to the $n~=1$ resonance interaction. Cyan vertical lines in panels (e) and (f) indicate $v_{\parallel}$ values at which $n=-1$ resonance is expected.} \label{fig.1}
\end{figure*}

\textcolor{black}{Figure~\ref{fig.1} depicts the the total electron VDF $f_e = f(v_x, v_y)$ at the initial stage of the simulation ($t_0=0$, Figure~\ref{fig.1}(a)), during the development of the oblique whistler heat-flux instability (Figures~\ref{fig.1}(b) and (c), respectively at $t_1=0.47\; \Omega_{ci}^{-1}$ and $t_2=0.94\; \Omega_{ci}^{-1}$), after the O-WHFI saturation (Figure~\ref{fig.1}(d)), during the development of the quasi-parallel WHFI (Figure~\ref{fig.1}(e)) and at the final stage ($t_{end}=5.6\; \Omega_{ci}^{-1}$) of the simulation (Figure~\ref{fig.1}(f)).
The strahl undergoes pitch-angle scattering (by the excited whistler waves, as will be shown below), which results in the reduction of the strahl drift velocity and in the simultaneous broadening of the strahl pitch angle distribution (Figure~\ref{fig.1}(b)). The scatter of the strahl electrons leads to the formation of a new population that can be seen as a halo, whose features are noticeable in the electron VDF from the onset of the O-WHFI to the final stages of the simulation (Figure~\ref{fig.1}(e)).} 

\begin{figure*}
 \centering
 \includegraphics[width=0.75\textwidth]{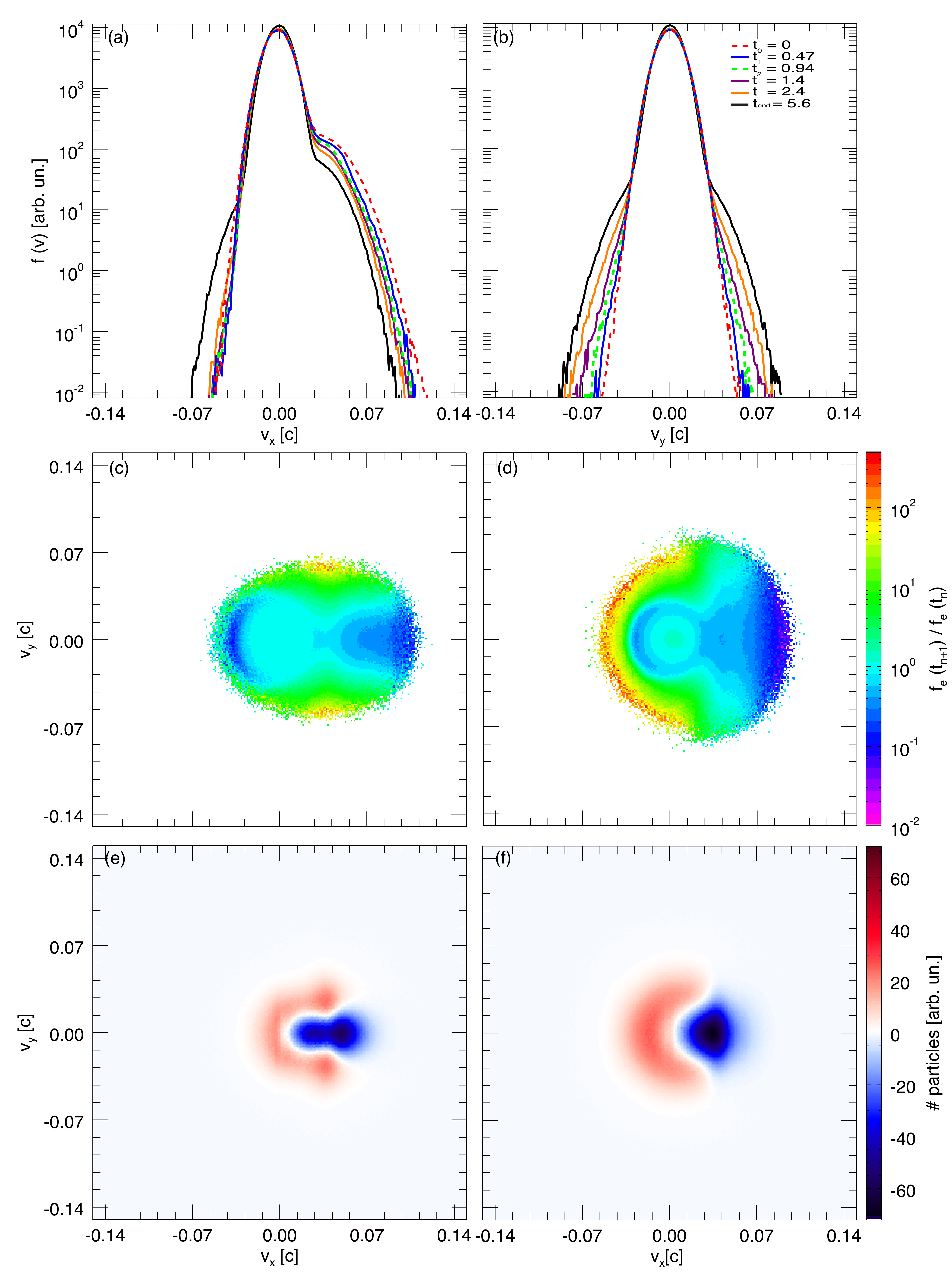}
 \caption{Total electron VDF cuts along the parallel ($f_e (v_x, v_y =0$, panel (a)) and perpendicular ($f_e (v_x =0, v_y$, panel (b)) directions. Colored lines indicate the cuts of the VDFs at different times. Temporal variation of the total electron VDF: $f_e (t_{2})\; /\; f_e (t_0)$ (panel (c)) and $f_e (t_{end})\; /\; f_e (t_2)$ (panel (d)). Variation of the strahl distribution function: $f_s (t_2) - f_s (t_0)$ (panel (e)) and $f_s (t_{end}) - f_s (t_2)$ (panel (f)).}\label{fig.2}
\end{figure*}

\textcolor{black}{To give a further evidence of the halo formation, in Figures~\ref{fig.2}(a) and (b) we compare the cuts of the total electron VDF along the parallel ($f_e (v_x, v_y =0)$) and perpendicular ($f_e (v_x =0, v_y)$) directions, for the times shown in Figure~\ref{fig.1}. One can observe the presence of the strahl component along the magnetic field direction, the relaxation of its drift velocity and the appearance of the suprathermal halo which, at higher energies, deviates from the Maxwellian distribution. In Figures~\ref{fig.2}(c) and (d) we show the temporal variation of the total electron VDF as can be seen from the ratio $f_e (t_{2})\; /\; f_e (t_0)$ and $f_e (t_{end})\; /\; f_e (t_2)$, respectively, while in Figures~\ref{fig.2}(e) and (f) we show the temporal variation of the strahl VDF: $f_s (t_{2}) - f_s (t_0)$ and $f_s (t_{end}) - f_s (t_2)$, respectively. All four panels confirm the generation of suprathermal halo particles distributed at all pitch angles at the expense of the strahl.}

\begin{figure*}
\centering
 \includegraphics[width=0.75\textwidth]{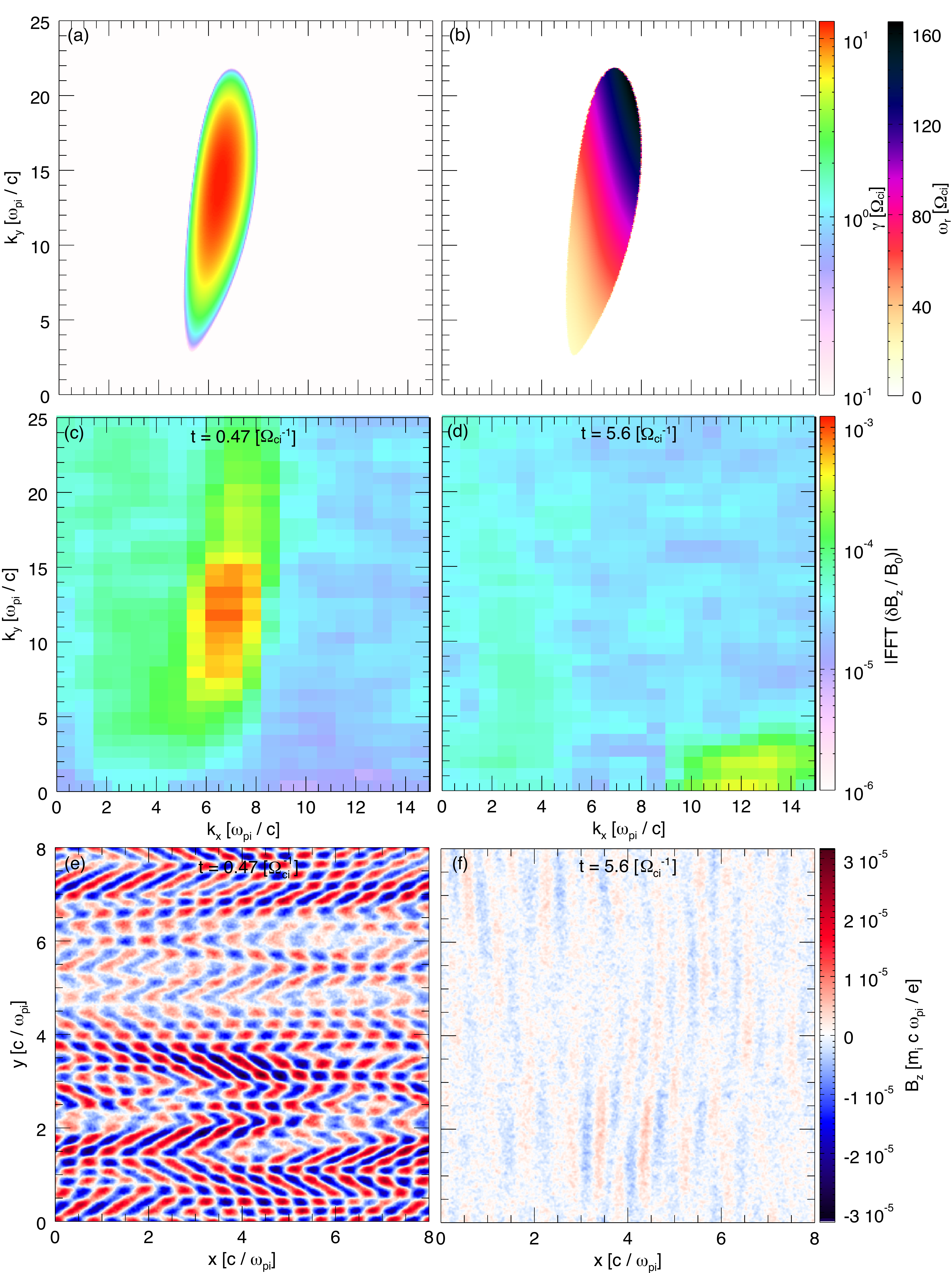}
\caption{Oblique Whistler Heat Flux Instability growth rates $\gamma$ (a) and wave frequencies $\omega_r$ (b) in the $k_x – k_y$ plane obtained from the linear dispersion relation. Fast Fourier transforms of the simulated transverse magnetic fluctuations $\text{FFT}(\delta B_z/B_0)$ at $t_{1} = 0.47\; \Omega_{ci}^{-1}$ (c) and $t_{end} =5.6\; \Omega_{ci}^{-1}$ (d). $B_z$ component (color scale) in the x-y plane at $t_{1} = 0.47\; \Omega_{ci}^{-1}$ (e) and $t_{end} =5.6\; \Omega_{ci}^{-1}$ (f).} \label{fig.3}
\end{figure*}

\textcolor{black}{In Figures~\ref{fig.3}(a) and (b) we report the growth rates $\gamma$ and the corresponding real wave frequencies $\omega_r$ in the $k_x – k_y$ plane, obtained from the theoretical linear dispersion relation. The strongest instability induced by the core-strahl electron VDF is the purely oblique right-handed O-WHFI, with maximum growth rate $\gamma_\text{max}\approx12.5\; \Omega_{ci}$ at $k_x\approx6.5\; \omega_{pi}/c$ and $k_y\approx13.5\; \omega_{pi}/c$. The unstable whistler modes have frequencies in the range of $\Omega_{ci} < \omega_r < \Omega_{ce}$ and present their maximum growth rate at $\theta \approx 65^{\circ}$.\\
The simulation clearly exhibits evidence of this instability. Indeed, the FFT of the simulated transverse magnetic fluctuations, normalized to the background magnetic field ($FFT(\delta B_z/B_0)$), at $t_1=0.47\; \Omega_{ci}^{-1}$ displayed in Figure~\ref{fig.3}(c), shows that the power is concentrated at highly oblique angles, $k_y\approx13 \; \omega_{pi}/c$ and between $k_x = 6$ and $8\; \omega_{pi}/c$.
At $t_{end}=5.6\; \Omega_{ci}^{-1}$, when the O-WHFI is already saturated, we observe weak modes which are mainly parallel or quasi-parallel to the background magnetic field, see Figure~\ref{fig.3}(d). These fluctuations concentrate at higher parallel wave numbers, between $k_x = 11$ and $13\; \omega_{pi}/c$, characteristic of the quasi-parallel WHFI. The transition from oblique to parallel modes is confirmed in Figures~\ref{fig.3}(e) and (f), where we show the transverse magnetic field in the $x – y$ plane at $t_{1} = 0.47\; \Omega_{ci}^{-1}$ and $t_{end} = 5.6\; \Omega_{ci}^{-1}$, respectively.} 

\begin{figure*}
 \centering
 \includegraphics[width=0.75\textwidth]{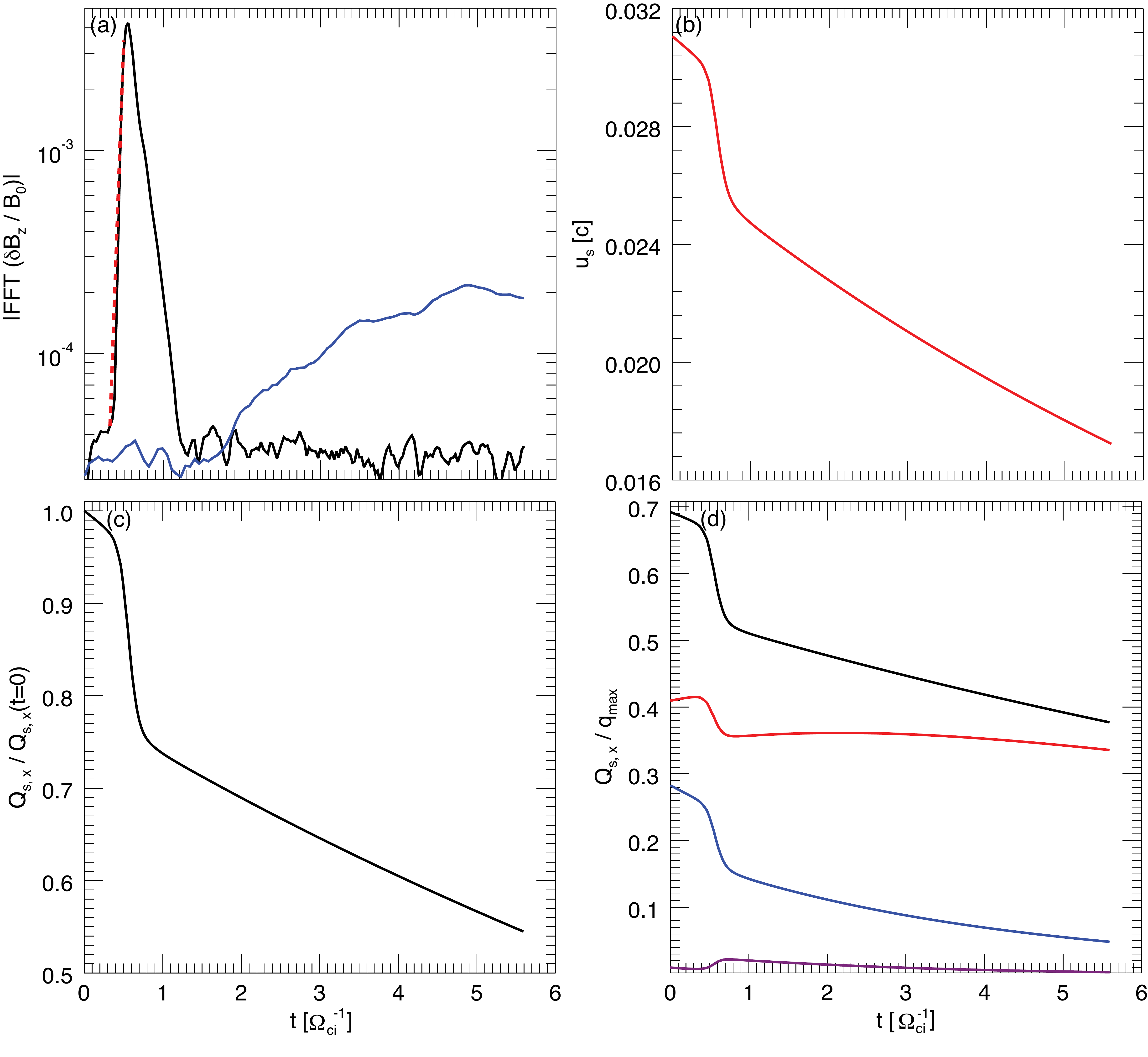}
 \caption{Temporal evolution of the simulated fastest growing oblique mode ($k_x = 6.5\; \omega_{pi}/c$ and $k_y = 13.5\; \omega_{pi}/c$, solid black line), of the simulated parallel mode at $k_x = 12\; \omega_{pi}/c$ and $k_y =0$ (solid blue line), and comparison with the maximum theoretical growth rate ($\gamma_\text{max} = 12.5\; \Omega_{ci}$, red dashed line) (a). Temporal evolution of parallel strahl drift velocity $u_s$ (b) and of strahl energy flux normalized to its initial value $Q_s / Q_s (t = 0)$ (c). Energy flux components carried by the electron strahl along the magnetic field direction: $Q_s$ (black), $Q_{\text{enth}, s}$ (red), $Q_{\text{bulk}, s}$ (blue), and $q_{s}$ (purple) (d). All the energy flux components are normalized to $q_{max} = 5/2 m_e (n_c w_c^3 + n_s w_s ^3)$.} \label{fig.4}
\end{figure*}

\textcolor{black}{The evolution of the fastest growing mode, at $k_x = 6.5\; \omega_{pi}/c$ and $k_y = 13.5\; \omega_{pi}/c$, obtained from the simulation, shows an exponential growth that saturates at $t\approx0.6\; \Omega_{ci}^{-1}$, followed by a relaxation, as shown in Figure~\ref{fig.4}(a). The maximum growth rate according to the linear theory is very similar to that obtained in our simulation. In Figure~\ref{fig.4}(a) we also show the temporal evolution of a parallel whistler mode selected through a cut of $\text{FFT}(\delta B_z/B_0)$ at $k_x = 12\; \omega_{pi}/c$ and $k_y =0$. Parallel and quasi-parallel modes are  manifested at later times and become dominant after $t\approx 2\; \Omega_{ci}^{-1}$.}

\textcolor{black}{The scattering of the strahl electrons by the generated whistler waves provokes the saturation of the O-WHFI instability but, at this point, the newly generated halo provides the energy for the excitation of quasi-parallel whistler waves (see Figure~\ref{fig.3}(d)). A further portion of suprathermal electrons is scattered by the enhanced quasi-parallel whistler fluctuations and the process results in a consequent relaxation of the strahl drift velocity, in the formation of a more symmetric halo, and leads to a more isotropic total electron VDF.} 

\textcolor{black}{To explore the physical nature of the strahl electron scattering, we investigate the resonance conditions. In Figures~\ref{fig.1}(b) and (c), the black and red vertical lines identify the parallel velocities at which the $n=0$ (Landau) and $n=1$ (cyclotron) resonant interactions of the fastest growing whistler wave with electrons are expected, calculated as $v_{\parallel} = (n~ \Omega_{ce} + \omega_{r}) /k_{\parallel}$, while circles show the electrons diffusion paths due to the $n~=1$ resonance interaction, as constant energy surfaces in the wave frame of reference \citep{Verscharen_2019}. The circles are centered at $v_{\parallel} = v_{ph}$ and $v_{\perp} =0$, with $v_{ph} = \omega_{r} /k_{\parallel}$ being the parallel phase velocity of the waves.  The $k_{\parallel}$, which in our simulation corresponds to $k_x$ of the fastest growing wave, is computed via fast Fourier transform (FFT) in space of the transverse magnetic fluctuations (see Figures~\ref{fig.3}(c) and (d)), while $\omega_r$ is obtained from the linear dispersion relation (see Figure~\ref{fig.3}(b)) for $k_{\parallel}$ and $k_{\perp}$ of the fastest growing wave. We consider the variation of $k_{\parallel}$ during the development of the O-WHFI and we take $k_{\parallel} = 6.5\; \omega_{pi}/c$ at $t_1=0.47\; \Omega_{ci}^{-1}$ and $k_{\parallel} = 8\; \omega_{pi}/c$ at $t_2=0.94\; \Omega_{ci}^{-1}$. The red vertical lines, respectively drawn at $v_{\parallel} =0.068  ~c$ in Figure~\ref{fig.1}(b) and $v_{\parallel} =0.057  ~c$ in Figure~\ref{fig.1}(c), delineate the electron population that fulfills the $n=1$ resonance condition and diffuses, via pitch-angle scattering, towards higher values of $v_{\perp}$. Indeed, in both panels, the circles closely correspond to the non-Maxwellian horn-like structures of the electron VDF (around $v_x = 0.05 ~c$ at $t=t_1$ and around $v_x = 0.035 ~c$ at $t=t_2$). The Landau resonance ($n=0$) is not prominent in the $v_{\parallel} > 0$ region of the VDF in the early stage of the simulation as $\partial f_e / \partial v_{\parallel} < 0$. However, on the left side of horn-like extensions we see $\partial f_e / \partial v_{\parallel} > 0$, so the Landau resonance starts to be effective when the horns reach the black vertical line (around $t_2 = 0.94\; \Omega_{ci}^{-1}$). The Landau resonance scatters electrons along $v_{\parallel}$, contributing to the isotropization of the scattered electrons and the formation of the electron halo.  
The generation of the tail in the distribution function at $v_{\parallel} < 0$ (see also Figures~\ref{fig.2}(a) and (d)) can be traced back to the $n=-1$ cyclotron resonance \citep{Verscharen_2019,Roberg-Clark2019} that the electron VDF experiences later in time, when the there is a sufficient number of electrons with $v_x < 0$ and quasi-parallel WHFI is triggered. Particles with high $v_{\perp}$ that fulfil this resonant condition are scattered by waves towards lower values of $v_x$ (in the direction opposite to $B_0$), leading to the formation of a more symmetric halo. In Figures~\ref{fig.1}(e) and (f) the cyan vertical lines indicate $v_{\parallel}$ values at which $n=-1$ resonance interactions of electrons with quasi-parallel whistler waves are expected. The $\omega_r$ values, at these stages, are obtained by solving the linear dispersion relation for the parallel whistler instability triggered by isotropic drifting suprathermal population of electrons.}

In Figures~\ref{fig.4}(b), (c) and (d) we \textcolor{black}{display} the temporal evolution of the strahl drift velocity $u_s$ and the heat flux $Q_s$ carried by the strahl along the magnetic field direction.
Since the drift velocities of the core and the strahl are initialized in the parallel direction with respect to the magnetic field, the parallel component of the heat flux carried by the strahl in the reference frame of the solar wind (strahl energy flux) is dominant compared to the perpendicular components. It is determined by $Q_{s} = \frac{m_s}{2} \int v_{\parallel} v^2 f_s d^3v$ and it can be broken into several components \citep{Feldman1975,Innocenti2020} as $Q_{s} = Q_{\text{enth}, s} + Q_{\text{bulk}, s} + q_s$,  where $Q_{\text{enth}, s} = \frac{3}{2}~ n_s m_s u_s w_s^2 $ reflects the convection of the strahl electron enthalpy, $Q_{\text{bulk}, s} = \frac{1}{2}~ m_s n_s u_s^3$ is the energy flux due to the bulk motion of the electrons, and $q_{s} = \frac{m_s}{2} \int (v_{\parallel} - u_s) (v - u_s)^2 f_s d^3v$ represents the heat flux carried by the strahl in its reference frame (skewness of the strahl VDF).
The decrease of the strahl energy flux is \textcolor{black}{around 46$\%$ during the entire simulation}. Comparison of Figure~\ref{fig.4}(a) and Figures~\ref{fig.4}(b) and (c) shows that the strongest heat flux rate decrease is simultaneous with the growth of the oblique wave modes and the consequent decrease of the strahl drift velocity that it produces, and it lasts until their saturation. After the O-WHFI is already saturated, the plasma is still unstable to the quasi-parallel WHFI (as shown in Figure~\ref{fig.3}) and then the heat flux is subject to a further decrease. It is important to note that the rate of reduction of the heat flux, \textcolor{black}{due to the O-WHFI (before $t \approx 0.8\; \Omega_{ci}^{-1}$)}, is almost one order of magnitude higher than \textcolor{black}{the rate of the heat flux reduction due to quasi-parallel wave modes}.  
The heat flux is mostly carried by the convection of the strahl electron enthalpy, indicated in Figures~\ref{fig.4}(d) with a red line, in agreement with the observations reported by \citet{Feldman1975}. The regulation of the global strahl energy flux $Q_s$ along the background magnetic field is essentially produced by the relaxation of the strahl drift velocity rather than by the variation of the thermal velocity $w_j$ similarly to the results reported in \citet{Innocenti2020}.
The heat flux carried by the strahl in its reference frame represents a small part of the total energy flux. However it exhibits an increase when oblique whistler waves interact with the electron VDF and then a reduction due to the scatter of the strahl trough lower drift velocities, as $q_{s}$ is a direct measure of the electron VDF deformation.

\section{Discussion and Conclusions}\label{sec.5}
\textcolor{black}{We perform a 2D fully kinetic simulation to investigate the role of the oblique and parallel branches of whistler heat flux instability in shaping the electron VDFs in the solar wind. We confirm that, in a plasma consisting of a drifting core and strahl electrons, as recently observed by PSP in the near-Sun solar wind, whistler waves, propagating at oblique angles with respect to the background magnetic field, can be excited, in agreement with \citet{Verscharen_2019} and \citet{Lopez2020}. The free energy of the counter-streaming populations of electrons is converted into magnetic energy in the form of oblique whistler-mode waves that produce an enhanced and rapid suppression of the plasma heat flux until their saturation. The oblique whistler waves also drive a significant pitch-angle scattering of the field-aligned strahl, which results in the formation of a suprathermal electron halo population. A portion of the strahl population of electrons, whose parallel velocities satisfy the $n=1$ cyclotron resonance condition, is scattered towards high values of perpendicular velocities, producing strong deviations of the VDF from the original distribution. Our non-linear study allows us to conclude that the excited whistler-mode waves shift towards smaller angles of propagation as the bulk velocity of the strahl decreases and that the diffusion of the electrons to higher $v_{\perp}$ occurs mostly during the linear and non-linear stages of the oblique whistler heat flux instability. During the non-linear stage, the scattering of electrons in the parallel direction starts, at least partly due to the $n=0$ Landau resonance with the oblique whistler waves. Later on, when there is a sufficient number of suprathermal electrons with $v_x <0$, the electron system experiences secondary effects due to the $n=-1$ cyclotron resonant interaction with parallel whistler waves. Even if the parallel modes saturate at moderate levels of magnetic field fluctuations and reduce the heat flux only slowly, they do lead to a further relaxation of the suprathermal electrons, to the generation of a tail-like structure in the distribution function at $v_x < 0$ and hence to a more symmetric halo, in agreement with observations.}

We notice that while modelling the initial distribution function of the strahl population as a \textcolor{black}{drifting bi-}Maxwellian provides an accurate representation for pristine solar wind conditions \citep{Halekas2020, Bercic2020}, it may not be accurate at large heliocentric distances. In this case, simulations should be initialized with a different distribution function \citep[e.g.][]{Horaites2018}. The study of the WHFI for non-Maxwellian strahl VDF is more relevant close to the Earth and \textcolor{black}{is beyond the scope of this work.}

Future in situ observations by PSP at even smaller heliocentric distances may further clarify the physical processes affecting the electrons, since in the close vicinity of the Sun, before the propagation through the heliosphere, electron VDFs are less affected by different micro-instabilities. This may give us an opportunity to detect pristine electron VDFs and thus gain an insight into the processes that influence solar wind particle distributions at the early stages of their evolution.    

\acknowledgments
Authors thank J. S. Halekas for helpful discussions.
This work was supported by a PhD grant awarded by the Royal Observatory of Belgium to one of the authors (A. M.).
These simulations were performed on the supercomputers SuperMUC (LRZ) and Marconi (CINECA) under PRACE allocations.
A. N. Z. thanks the European Space Agengy (ESA) and the Belgian Federal Science Policy Office (BELSPO) for their support in the framework of the PRODEX Programme.
R.A.L thanks the support of AFOSR grant FA9550-19-1-0384.
M.E.I.'s work is supported by an FWO postdoctoral fellow.

\bibliographystyle{aasjournal}  
\bibliography{main}

\end{document}